\documentclass[%
reprint,
superscriptaddress,
amsmath,amssymb,
aps,
floatfix,
]{revtex4-2}

\usepackage{graphicx}% Include figure files
\usepackage{dcolumn}% Align table columns on decimal point
\usepackage{bm}% bold math
\usepackage{balance}
\usepackage{graphicx}
\usepackage{dcolumn}% Align table columns on decimal point
\usepackage{siunitx}
\usepackage{xr}
\usepackage{xcolor}
\usepackage[utf8x]{inputenc}

\makeatletter

\begin{document}

% \preprint{}

\author{S. Arora}
\affiliation{Kavli Institute of Nanoscience, Delft University of Technology, 2600 GA, Delft, The Netherlands}
\author{T. Bauer}
\affiliation{Kavli Institute of Nanoscience, Delft University of Technology, 2600 GA, Delft, The Netherlands}
\author{N. Parappurath}
\affiliation{Kavli Institute of Nanoscience, Delft University of Technology, 2600 GA,
Delft, The Netherlands}
\affiliation{Center for Nanophotonics, AMOLF, Science Park 104, 1098 XG Amsterdam, The Netherlands}
\author{R. Barczyk}
\affiliation{Center for Nanophotonics, AMOLF, Science Park 104, 1098 XG Amsterdam, The Netherlands}
\author{E. Verhagen}
\affiliation{Center for Nanophotonics, AMOLF, Science Park 104, 1098 XG Amsterdam, The Netherlands}
\author{L. Kuipers}\email{l.kuipers@tudelft.nl}
\affiliation{Kavli Institute of Nanoscience, Delft University of Technology, 2600 GA, Delft, The Netherlands}

\title{Breakdown of spin-to-helicity locking at the nanoscale in topological photonic crystal edge states}

\date{\today}
%TC:ignore

\begin{abstract} 
%\textcolor{cyan}{[$<$600 characters - currently 904]}

We measure the local near-field spin in topological edge state waveguides that emulate the quantum spin Hall effect.  We reveal a highly structured spin density distribution that is not linked to a unique pseudospin value.  From experimental near-field real-space maps and numerical calculations, we confirm that this local structure is essential in understanding the properties of optical edge states and light-matter interactions. The global spin is reduced by a factor of 30 in the near field and, for certain frequencies, flipped compared to the pseudospin measured in the far field.  We experimentally reveal the influence of higher-order Bloch harmonics in spin inhomogeneity, leading to a breakdown in the coupling between local helicity and global spin.
\end{abstract}

\maketitle
%TC:endignore

%\textcolor{cyan}{[$<$3750 words - currently 3040]}

Robust unidirectional transport of photonic states with coherent and highly efficient light-matter coupling is a luring proposition for large-scale on-chip quantum networks \cite{Borregaard2019,Pichler2016}. Unidirectional photon emission is achieved, in general, by strong coupling of a quantum emitter supporting a circularly polarized dipole moment to photonic states with a corresponding local polarisation orientation (helicity) \cite{Coles2016,Sollner2015,Young2015,Bliokh2015}. This results in a chiral quantum optical interface owing to spin-momentum locking \cite{VanMechelen2016}. However, many practical realizations of such chiral interfaces exhibit only a finite spin-to-direction coupling. These realizations are susceptible to disorder and defects \cite{Orazbayev2019}. More so, unambiguous spin-dependent transport of quantum information from chiral emitters requires an explicit account of the local structure of the edge state's optical spin density profile to achieve high directional coupling \cite{Proctor2019, Young2015, Deng2018, Wang2020, Mazor2020, Petersen2014,Proctor2020, Oh:18, Baranov2017, JalaliMehrabad2020, Sapienza2015}.

One class of systems that has been proposed in the context of robust spin-dependent transport, is the photonic crystal-based analog of topological insulators which emulate the quantum spin Hall effect (QSHE) \cite{hasan2010colloquium}. They feature two symmetry-protected edge states at the interface between a topologically trivial and non-trivial lattice \cite{Wu2015,Jacobs}. Experimental as well as numerical realizations have shown that the two edge states each exhibit a unique pseudospin due to the different topological invariants of the supporting bulk bands, that makes photonic transport robust to defects and sharp corners \cite{Barik2018, Pregnolato2020, JalaliMehrabad2020}. This transport relies on pseudospin coupling to the far-field (FF) helicity to ensure and maximize photon unidirectionality \cite{Christiansen2019}. Leveraging the helicity supported by these systems has enriched applications for quantum entanglement \cite{Deng2018} and quantum spin circuits \cite{Pichler2015, Mahmoodian2016, St-Jean2017, Hafezi2011, Rider2019}. However, it is essential to determine the exact relation between field helicity and edge state pseudospin.

In this paper, we examine the near field of edge states in topological photonic crystals (TPCs) to comprehensively study the (local) chiral information. With phase- and polarisation-resolved near-field optical microscopy (NSOM), we collect the orthogonal in-plane polarisation components of the electric field and determine the underlying spatially varying spin density. We experimentally verify that the inhomogeneity in optical spin density follows the Bloch periodicity of the lattice. By experimentally accessing the different Bloch harmonics that together form the symmetry-protected edge state, we show that accounting for the individual contribution of each higher-order Bloch harmonic breaks the coupling between pseudospin and helicity of the edge state.

\begin{figure}[hbt]
	\centering
    \includegraphics[]{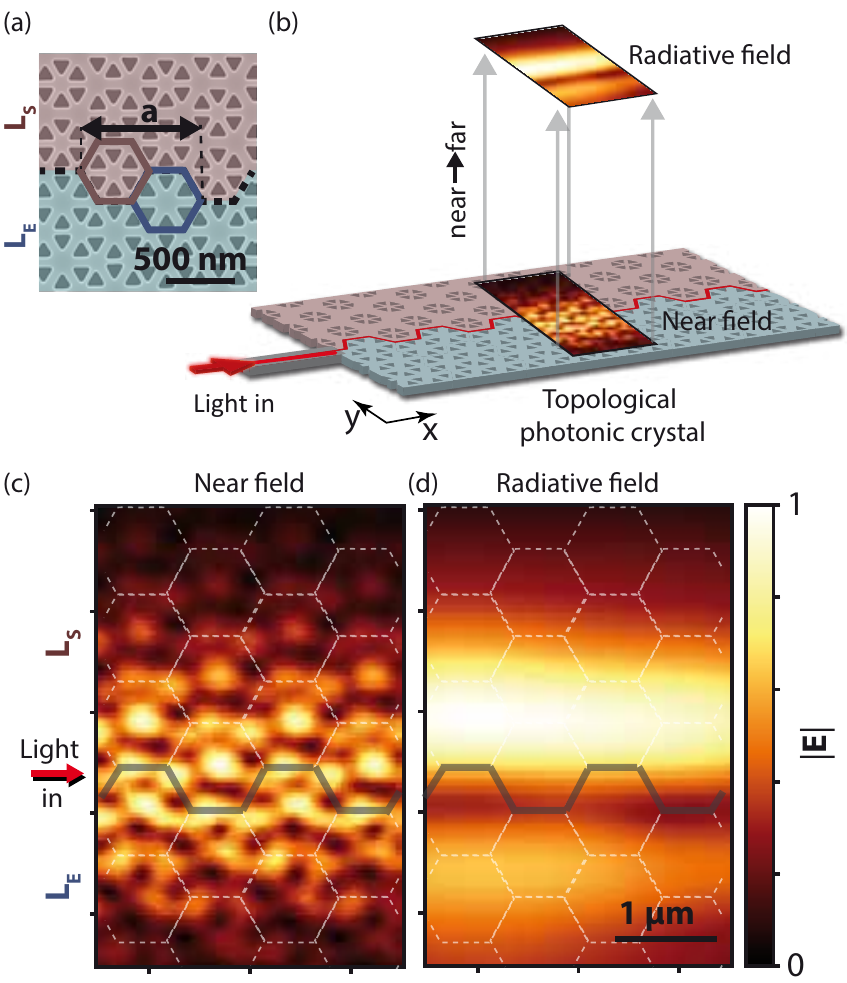}
	\caption{(a) Scanning electron micrograph (SEM) of the topological interface in the fabricated sample with the color-coded regions depicting the shrunken $L_S$ (red) and expanded $L_E$ (blue) lattices. The lattice periodicity is $a=\SI{880}{nm}$. (b) Schematic representation of the TPC lattice with overlaid near- and far-field amplitudes of the electromagnetic field. Close-up of the normalized in-plane electric (c) near-field and (d) radiative field amplitude in the TPC featuring an armchair interface at $\lambda =  \SI{1520}{nm}$ with the light launched into the structure from the left (indicated by the red arrow). The dashed hexagonal pattern outlines the underlying crystal lattice.}
	\label{fig1}
\end{figure}

Following the shrink-and-grow design \cite{Wu2015, Jacobs, Parappurath2020}, we realize a TPC interface on a silicon-on-insulator platform by deforming a graphene-like hexagonal lattice with six equilateral triangular holes. While the unperturbed lattice features a doubly degenerate Dirac cone at the $\Gamma$-point in the dispersion diagram, this degeneracy is lifted in two ways (see Fig. \ref{fig1}(a)): on one side of the interface, the holes are concentrically shifted inwards, called the shrunken lattice (L$_S$), while on the other side, the holes are shifted concentrically outwards, labelled as expanded lattice (L$_E$). The geometrical transformations keep the global C$_6$ symmetry of each lattice unaltered. The band structure of L$_S$ and L$_E$ both reveal a direct bandgap at the $\Gamma$-point. For L$_S$, the shape of the electric field in the lower band resembles `p'-like orbitals and the upper band resembles `d'-like orbitals, whereas for L$_E$, the mode symmetries of the upper and lower band is inverted \cite{Wu2015}. The different intra- and inter-cell coupling strengths between neighboring sites in the two lattices imparts a non-trivial nature to the expanded lattice \cite{Asboth2016}. The interface of L$_S$ and L$_E$ supports two counter-propagating edge states that traverse the band gap around the $\Gamma$-point \cite{Wu2015, Jacobs}. These edge states are robust against back-scattering and offer unidirectional transport, provided no scattering between pseudospins occurs. The TPC reported in this paper is different from a photonic crystal emulating the quantum valley Hall effect \cite{Arora2021, Shalaev2018a, He2019, Tzuhsuan2016, Dong2017}, as the latter supports edge states that lie around the high-symmetry points $K$ and $K^\prime$ of the Brillouin zone. Since the QSHE TPC edge states traverse the $\Gamma$-point, the counter-propagating modes lie above the light cone and therefore couple to FF radiation. This makes them accessible to far-field spectroscopic investigations \cite{Parappurath2020, Gorlach2018, Barik2018}. Here, the polarization of the light scattered to the FF shows a near-unity optical spin and can be directly linked to the state's pseudospin \cite{Parappurath2020, Yang2018, ZhengWangY2008}. However, the radiative FF does not contain the full complex information of the evanescent electromagnetic field since it only takes into account the plane waves within the light cone \cite{Kaspar2011, busch2006photonic} and a detailed account of the full local field is imperative for ensuring chiral light-matter interactions on the nanoscale.

\begin{figure}[t]
	\centering
    \includegraphics[]{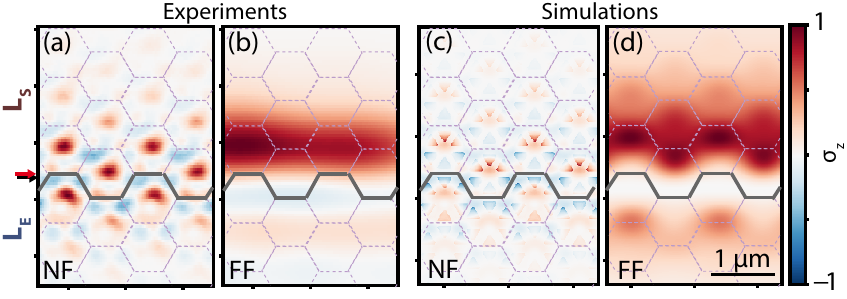}
	\caption{(a) Experimentally measured NF spin density $\sigma_z$ of the AC-edge mode over two unit cells for an excitation wavelength of $\lambda=\SI{1520}{nm}$. (b) Experimentally measured FF spin density over the same extent as in (a), realized by filtering out the non-radiative wavevectors of the field shown in (a). (c) Numerically calculated spin density for $k_x = \SI{0.08}{(\pi/a)}$ with (d) displaying the spin density for numerical simulations in the FF. The solid gray line indicates the armchair interface while the dashed hexagonal pattern outlines the underlying crystal lattice.}
	\label{fig2}
\end{figure} 

To experimentally investigate the spin character of the TPC interface rigorously, we fabricate a lattice featuring an armchair (AC) interface. Fig. \ref{fig1}(a) shows the unit cell structure on both sides of the interface. We measure the complex in-plane electric field distribution E using a phase- and polarization-resolving near-field scanning optical microscope \cite{Rotenberg2014}. The light propagation along the interface is schematically visualized in Fig. \ref{fig1}(b). Here, we distinguish the two typical evaluation regimes of the electromagnetic (EM) field: near (NF) and radiative (FF). The former defines the component of the EM field directly above the crystal surface, constituting the decaying evanescent field as well as the radiative field ($\textbf{E}_\textbf{rad}$) of the propagating edge state. The latter typically forms at distances of multiple wavelengths away from the interface and is associated with only $\textbf{E}_\textbf{rad}$. In Fig. \ref{fig1}(c), we depict the measured electric field amplitude at an excitation wavelength of $\lambda = \SI{1520}{nm}$. A highly structured field is visible. The mode amplitude diminishes while propagating from left to right (zoom-in section shown in Fig. \ref{fig1}(c)), associated with radiation leaking to the FF \cite{Parappurath2020, Yang2018, ZhengWangY2008}. The field profile around the AC interface (indicated by the solid gray line in Fig. \ref{fig1}c) resembles a Bloch wave \cite{Russell1986} with pockets of high and low field amplitude forming a hexagonal pattern that repeats with lattice periodicity $a$ along the propagation direction $+x$. The dominant part of the field is confined to the interface. The transverse extent of the edge state shows that the field extends further into the L$_S$ bulk than in L$_E$ (see Supplementary Fig. S1 for more information on longitudinal and transverse extent). This asymmetry in evanescent tails is consistent with the relative width of the photonic bandgap of the lattices, which is larger for L$_E$. The structured NF information is transformed to the radiative FF of the edge state by limiting the collected wavevector range to lie within the light cone using Fourier filtering. (see white dashed circle in reciprocal space in Fig. \ref{fig4}). In contrast to Fig. \ref{fig1}(c), where the amplitude antinodes follow the underlying crystal structure, the FF amplitude is homogeneous along the TPC interface (see Fig. \ref{fig1}(d)). This is consistent with the restriction in $k$-space resulting in a single in-plane $k$-vector associated to the mode of the edge state in the direction of propagation. Predictably, the dominant field energy in the FF lies more on the shrunken side of the interface, consistent with NF observations and the corresponding size of the bulk band gaps. An evanescent field strongly confined to the interface, propagating in the $+x$ direction and decaying in $y$-direction, implies the existence of transverse spin \cite{Bliokh2015a, Aiello2015a, Bliokh2015c, Bekshaev2015} with positive and negative helicity mirrored at the interface.  For our C$_6$ symmetric lattice, this results in the dominance of one designated helicity that is locked to the direction of mode propagation. The calculated transverse optical spin of the evanescent field is therefore a non-zero value. However, the spin distribution of the mode in the NF and FF differs.

We calculate the spin density distribution $\sigma_z (x,y) $ =  $\mathrm{Im}$ [\textbf{E$^*(x,y)$} $\times$ \textbf{E$(x,y)$}]$_z$ \cite{Aiello2015a}, where \textbf{E$^*(x,y)$} is the complex conjugate of the electric NF. Fig. \ref{fig2} depicts the measured and calculated $\sigma_z (x,y)$. For the evanescent field, $\sigma_z$ is the expectation value of the helicity of light and it directly translates to the \textit{local} field polarization state of the photonic TE-like mode as a result of spin-orbit interactions. The NF $\sigma_z$ depicted in Fig. \ref{fig2}(a) reveals a highly structured $\sigma_z$ distribution. A periodic pattern of $+\sigma_z$ and $-\sigma_z$ is observed, that repeats with a periodicity of $a$ in the propagation direction and $a/\sqrt{3}$ in the transverse direction. Close to the center of any given unit cell indicated by dashed lines in Fig. \ref{fig2}(a), antinodes of $+\sigma_z$ are prominent, whereas around the outlines of the unit cell, the sign of $\sigma_z$ flips. This spin flip within a unit cell confirms that the local handedness of a topological edge mode's polarization state is non-uniform. The local inhomogeneity of the spin density $\sigma_z (x,y)$ in the NF

\begin{equation*}
    \textbf{E}_\mathrm{\textbf{NF}}(x,y) = \iint_{-\infty}^{\infty} \widetilde{\textbf{E}}(k_x,k_y)e^{i[k_xx + k_yy]} \,dk_x \,dk_y ,
\end{equation*} 

which takes into account all spatial frequencies, completely vanishes in the FF

\begin{equation*}
    \textbf{E}_\mathrm{\textbf{rad}}(x,y) = \iint_{k_x^2 + k_y^2 < k_0} \widetilde{\textbf{E}}(k_x,k_y) e^{i[k_xx + k_yy]} \,dk_x \,dk_y
\end{equation*}
(shown in Fig. \ref{fig2}(b)), where $k_0 = \omega/c$. For an excitation wavelength of $\lambda = \SI{1520}{nm}$, only $+\sigma_z$ is visible close to the interface. This is in perfect agreement with reports of near-unity spin density in FF measurements \cite{Parappurath2020}. Comparing both regimes to numerical simulations in Fig. \ref{fig2}(c) and (d) reveals an excellent agreement of the $\sigma_z$ distribution for the NF as well as the FF case, respectively. To understand the origin of the observed spatial variations and difference in NF and FF, we undertake a detailed analysis of the edge states in momentum space. 

\begin{figure}
	\centering
	\includegraphics[width=\linewidth]{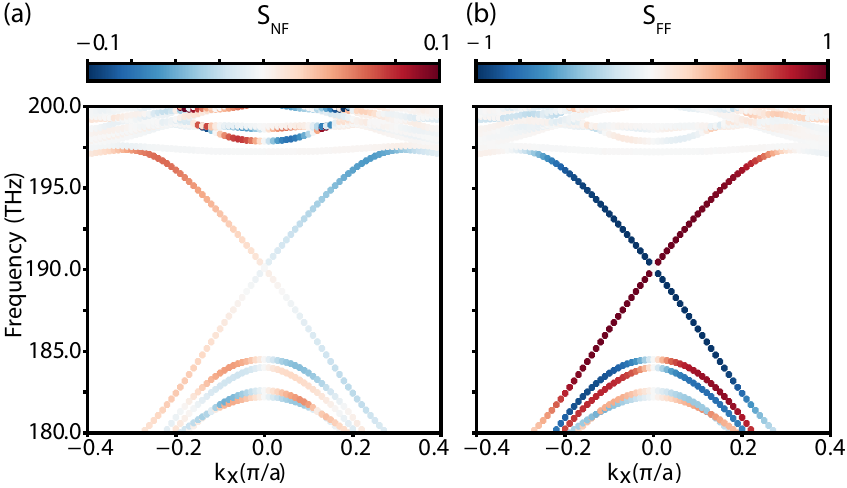}
	\caption{Numerically simulated dispersion relation where the edge state eigenfrequencies are color coded with the estimated (a) near-field and (b) far-field optical spin $S$.}
	\label{fig3}
\end{figure}
 
\begin{figure}[ht]
	\centering
	\includegraphics[width=\linewidth]{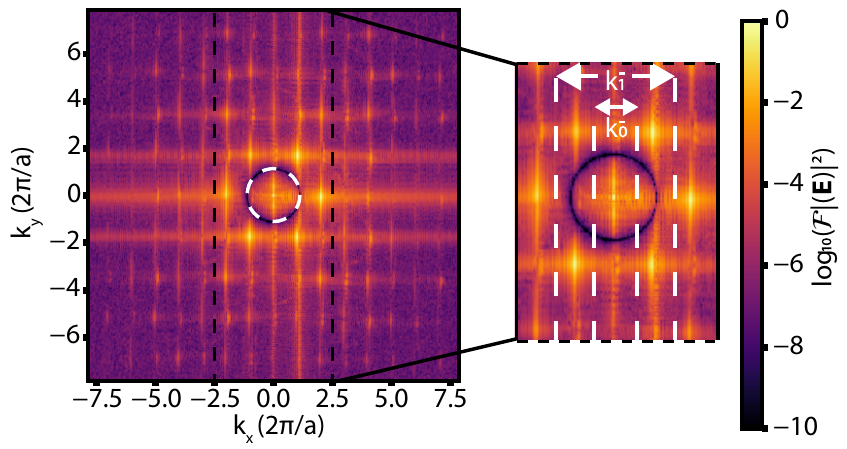}
	\caption{Two-dimensional Fourier representation of the experimentally measured electric field amplitude of the TPC edge mode at a wavelength of $\lambda = \SI{1560}{nm}$, with the amplitude shown in logarithmic scale. The reciprocal lattice vectors are periodically separated in the propagation direction ($k_x$) in units of $2 \pi/a$, where $a$ is the lattice constant of the armchair edge. The white dashed circle represents the light cone in air. Bottom inset: Zoom-in of momentum space restricted to show the fundamental ($k_{\bar{0}}$) and first higher-order BH ($k_{\bar{1}}$).}
	\label{fig4}
\end{figure}

We investigate the dissimilarity in the NF and FF optical spin density distribution by analysing the \textit{global} optical spin $S = \iint \sigma_z(x,y) \,dx\,dy$ of the edge state over its full band dispersion, representing the integrated helicity of the interface mode. Fig. \ref{fig3} shows the calculated eigenmodes of the AC interface for both regimes. Figures \ref{fig3}(a) and (b) both show a linear dispersion for the edge states that lie within the band gap and cross at the $\Gamma-$point. The edge states are disconnected from the top and bottom bulk bands and recombine at the Brillouin zone edge. We notice an anti-crossing of $\SI{0.1}{THz}$, predicted by the extended tight-binding model \cite{Chen2015a}. This is a result of spin-spin scattering due to coupling of the counter-propagating edge states governed by the local $C_6$ symmetry breaking at the interface of the TPC. A surprising difference between Fig. \ref{fig3}(a) and (b) lie in the colors that represent the sign of the optical spin  $S$. Fig. \ref{fig3}(a) shows the dispersion obtained by calculating $S$ from the in-plane field distributions and is referred to as the NF optical spin ($S_\mathrm{NF}$). As expected, the degenerate counter-propagating edge states exhibit opposing helicity. However, we observe that this helicity is flipped in Fig. \ref{fig3}(b), where we plot the FF spin  ($S_\mathrm{FF}$) of the edge states. The linear state with negative $S_\mathrm{NF}$ possesses positive $S_\mathrm{FF}$ and vice versa. Moreover, the $S_\mathrm{FF}$ exhibits a near-unity value, more than an order of magnitude larger than the maximum $S_\mathrm{NF} = \pm 0.056$. The tight-binding approach dictates that the pseudospin for each edge state is uniquely linked to its FF helicity. On the other hand, the much lower $S_\mathrm{NF}$ and the spin flip suggest that the pseudospin of the full electromagnetic mode of the edge state is in fact not uniquely linked to a designated $S_\mathrm{NF}$. The principal difference between NF and FF observations stems from the fact that the evanescent NF contains information from all higher-order Bloch harmonics, which are unaccounted for in the FF. To better understand this intriguing spin-flip transition, we investigate the spin of the individual Bloch harmonics.

The origin of the spatially varying spin distribution of the Bloch periodic structure is confirmed by performing a two-dimensional Fourier transform $\mathcal{F}(k_x , k_y)$ of the measured complex field amplitude (shown for $\lambda=\SI{1560}{nm}$ in Fig. \ref{fig4}). The high-intensity peaks arranged in a hexagonal pattern in reciprocal space are the result of the underlying $C_6$ symmetry. We remove unrelated contributions of stray light grazing along the surface of the crystal by Fourier filtering the light cone (indicated as white dashed circle in Fig. \ref{fig4}). For the chosen wavelength, within this light cone lies a central peak at $k_y = 0$ that corresponds to the part of the Bloch mode radiating to the FF. Along $k_x$ at multiples of $2\pi/a$, clusters of peaks are seen that correspond to the different Bloch harmonics (BH) building up the intricate NF subwavelength structure of the edge state. An excellent signal-to-background ratio (S/B) of \SI{40}{dB} is obtained in our experiment that allows us to resolve seven different BHs. These BHs, each contributing with different weights, together form the edge state mode which obeys Bloch's theorem \cite{Gersen2005b}. Individual contributions of individual BH to $S$ hold utmost significance in understanding the inhomogeneity in the underlying spin structure. 

\begin{figure}[ht]
	\centering
    \includegraphics[width=\linewidth]{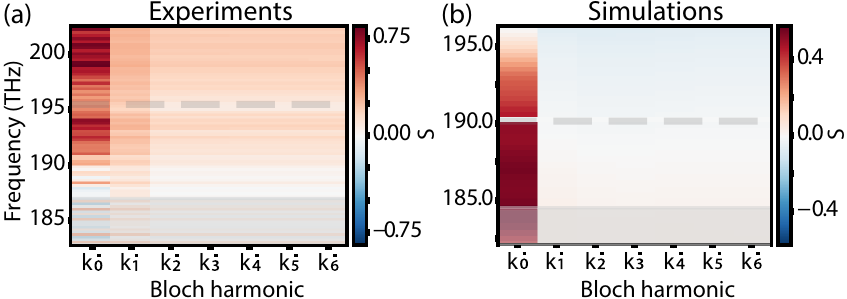}
	\caption{Build-up of the spin-dependent contributions of the individual BHs from (a) experiment and (b) numerical simulations, evaluated over several higher-order BHs. The gray dashed line indicates the frequency of the $\Gamma$-point crossing. The grayed out region extending below $\SI{184.4}{THz}$ depicts the lower band edge. }
	\label{fig5}
\end{figure}

We restrict our analysis in building up BH contributions by increasing the Fourier filter width in momentum space along $k_x$. In Figures \ref{fig5}(a) and (b), the first column represents the $S$ originating from the fundamental BH ($k_{\bar{0}}$) which extends from $k_x = [-0.5, 0.5]$, while the consecutive columns represent $S$ corresponding to all lowest-order BHs up and including the BH indicated in the horizontal axis label, i.e. the wavevector range $k_{\bar{n}}$ $\widehat{=}$ $[-1,1]*(n+0.5)$ (see right inset of Fig. \ref{fig4}). The quantitative analysis of the BH contribution to the optical spin is performed by first isolating BHs of certain width along $k_x$, applying an inverse Fourier transform to obtain the real-space field amplitude and subsequently calculating $S$. For $k_{\bar{0}}$, $S$ reflects the near-unity value that has been previously reported in FF measurements \cite{Parappurath2020}. Taking an exemplary frequency of \SI{188.92}{THz} in simulations, the radiative FF helicity $S_{k_{\bar{0}}}  = 0.53$. As the filter width in $k_x$ increases, the optical spin $S$ of the edge state reduces drastically for frequencies below the $\Gamma$-point crossing (at exemplary frequency the integrated NF spin $S_{k_{\bar{5}}} = -3.93 \num{1e-4}$). We acknowledge that the distinct difference between exact values of S for experiment and numerical calculation arise partly by the polarization sensitivity of the NF probe \cite{Gersen2007}. Nevertheless, $S_{k_{\bar{x}}}$ undergoes a clear reduction by an average factor of $30$ from $k_{\bar{0}}$ to $k_{\bar{1}}$, as shown in Fig. \ref{fig5}(b). This means that S cannot be a deterministic helicity parameter of the edge interface when including more than the fundamental BH since a one-on-one relation between local spin and helicity is no longer valid. The spin-momentum coupling breaks down to such an extent that it results in a sign flip as is evident in the sign switch around frequency $\SI{187.17}{THz}$ in Fig. \ref{fig5}(b). Therefore, we observe that already the first higher-order Bloch harmonic contribution to the \textit{global} optical spin completely breaks down pseudospin-momentum coupling.

In summary, we experimentally visualize the intricate spin density distribution of symmetry-protected edge states in topologically non-trivial photonic crystals that lie above the light line, using a NF microscope. We demonstrate that spin in such photonic systems no longer retains its unique handedness in comparison to electronic systems, where $\bm{s}$ must always be a good quantum number. We report that even the contribution of the first-order BH unambiguously flips the sign for certain excitation frequencies. Consequently, a \textit{priori} knowledge of detailed high spin density locations obtained from NF information will improve chances of precise positioning of quantum emitters along chiral interfaces. Without this knowledge, spin-polarised emission will result in mixing of pseudospin edge states, and therefore reduce the desired network efficiency. Thus, this finding needs to be accounted for in the architectures of future topological photonic quantum networks and it provides a pathway towards engineering truly robust topologically protected chiral interfaces.

The authors thank Filippo Alpeggiani and Aron Opheij for fruitful discussions about the initial design, fabrication and results. This work is part of the research programme of the Netherlands Organisation for Scientific Research (NWO). The authors acknowledge support from the European Research Council (ERC) Advanced Investigator Grant no. 340438-CONSTANS and ERC Starting Grant no. 759644-TOPP.

\bibliography{QSHE_bib}

\end{document}

%Title of paper
\title{Supplementary information: \\
Breakdown of spin-to-helicity locking at the nanoscale in topological photonic crystal edge states
% \\
% or
\\}

\author{S. Arora}
\affiliation{Kavli Institute of Nanoscience, Delft University of Technology, 2600 GA, Delft, The Netherlands}
\author{T. Bauer}
\affiliation{Kavli Institute of Nanoscience, Delft University of Technology, 2600 GA, Delft, The Netherlands}
\author{N. Parappurath}
\affiliation{Kavli Institute of Nanoscience, Delft University of Technology, 2600 GA,
Delft, The Netherlands}
\affiliation{Center for Nanophotonics, AMOLF, Science Park 104, 1098 XG Amsterdam, The Netherlands}
\author{R. Barczyk}
\affiliation{Center for Nanophotonics, AMOLF, Science Park 104, 1098 XG Amsterdam, The Netherlands}
% \author{F. Alpeggiani}
% \affiliation{Kavli Institute of Nanoscience, Delft University of Technology, 2600 GA, Delft, The Netherlands}
% \author{A. Opheij}
% \affiliation{Kavli Institute of Nanoscience, Delft University of Technology, 2600 GA, Delft, The Netherlands}
\author{E. Verhagen}
\affiliation{Center for Nanophotonics, AMOLF, Science Park 104, 1098 XG Amsterdam, The Netherlands}
\author{L. Kuipers}
\email{l.kuipers@tudelft.nl}
\affiliation{Kavli Institute of Nanoscience, Delft University of Technology, 2600 GA, Delft, The Netherlands}
\maketitle
\onecolumngrid
%\begin{center}
%Number of pages: x\\
%Number of figures: x
%\end{center}

%\newpage
% \section*{Glossary}
% \noindent
% TPC --- topological photonic crystal\\
% TMM --- transfer matrix model\\
%

\section{Near-field map of edge state}
\begin{figure}[hbt]
	\centering
    \includegraphics[width=0.75\linewidth]{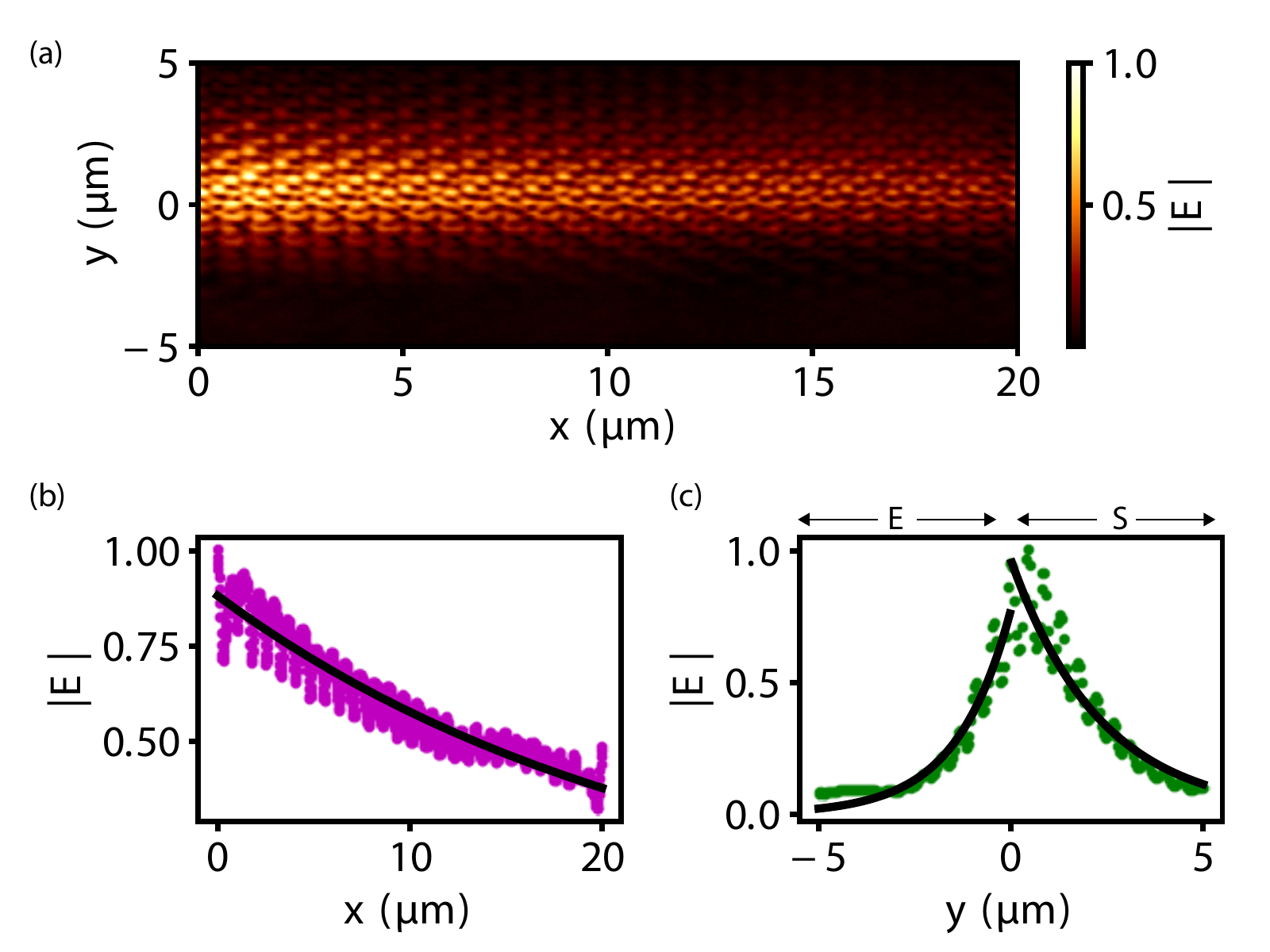}
	\caption{(a) Near-field map of a photonic crystal with an armchair edge: a. Experimentally measured electric field amplitude ca. \SI{20}{nm} above the surface of the photonic crystal at an excitation wavelength $\lambda = \SI{1560}{nm}$.
	b. Normalized integrated amplitude over the longitudinal extent of the edge state in the direction of propagation along the interface. The solid line corresponds to an exponential decay fit to the data. c. Normalized integrated amplitude from the transverse area. The two solid black lines indicate the exponential decay fit away from the interface into the bulk of the lattices (expanded on the left and shrunken on the right)}
	\label{fig1:supp}
\end{figure} 

Fig. \ref{fig1:supp}a shows the real space scan of the electric field amplitude at an exemplary wavelength of $\lambda = \SI{1560}{nm}$. The mode is confined to the armchair interface and decays exponentially along the propagation direction (see Fig. \ref{fig1:supp}b). We fit an exponential function to extract a decay length of $L = \SI{23.59}{\mu m}$. Fig. \ref{fig1:supp}c shows the transverse extent of the edge state away from the armchair interface into the expanded and shrunken lattices. The extracted decay length in the expanded lattice is $D_E = \SI{1.404}{\mu m}$ whereas the extracted decay length in the shrunken lattice is $D_S = \SI{2.35}{\mu m}$, which is consistent with the relative width of the photonic bandgap as discussed in the main text.

% \section{Fourier filtering}

% \bibliography{supplementary}